\newcommand{\CLASSINPUTbaselinestretch}{1.6}
\documentclass[journal,12pt,onecolumn,draftclsnofoot,]{IEEEtran}
\IEEEoverridecommandlockouts
\usepackage{cite}
\usepackage{amsmath,amssymb,amsfonts}
\usepackage{algorithmic}
\usepackage{float}

\usepackage{graphicx}
\usepackage{caption}
\usepackage{textcomp}
\usepackage{xcolor}
\usepackage{tabto}
\usepackage{wrapfig}
\usepackage{lscape}
\usepackage{rotating}
\usepackage{epstopdf}
\usepackage{pdfpages}
\def\BibTeX{{\rm B\kern-.05em{\sc i\kern-.025em b}\kern-.08em
    T\kern-.1667em\lower.7ex\hbox{E}\kern-.125emX}}

\begin{document}

\title{AEVB-Comm: An Intelligent Communication System based on AEVBs
}
\author{\IEEEauthorblockN{Raghu Vamshi Hemadri\IEEEauthorrefmark{1}, Akshay Rayaluru\IEEEauthorrefmark{1}, and Rahul Jashvantbhai Pandya\IEEEauthorrefmark{2}}
\IEEEauthorblockA{Dept. of Electronics and Communication Engineering,
National Institute of Technology, Warangal, India\\
Email: \IEEEauthorrefmark{1}vamshi.hemadri@gmail.com, \IEEEauthorrefmark{1}akshaynaidappa2@gmail.com, \IEEEauthorrefmark{2}rpandya@nitw.ac.in}}

\maketitle

\begin{abstract}
In recent years, applying Deep Learning (DL) techniques emerged as a common practice in the communication system, demonstrating promising results. The present paper proposes a new Convolutional Neural Network (CNN) based Variational Autoencoder (VAE) communication system. The VAE (continuous latent space) based communication systems confer unprecedented improvement in the system performance compared to AE (distributed latent space) and other traditional methods. We have introduced an adjustable hyperparameter $\beta$ in the proposed VAE, which is also known as $\beta-$ VAE, resulting in extremely disentangled latent space representation. Furthermore, a higher-dimensional representation of latent space is employed, such as 4n dimension instead of 2n, reducing the Block Error Rate (BLER). The proposed system can operate under Additive Wide Gaussian Noise (AWGN) and Rayleigh fading channels. The CNN based VAE architecture performs the encoding and modulation at the transmitter, whereas decoding and demodulation at the receiver. Finally, to prove that a continuous latent space-based system designated VAE performs better than the other, various simulation results supporting the same has been conferred under normal and noisy conditions.  
\end{abstract}

\footnote{IEEE-copyrighted material:© 2020 IEEE (paper is submitted for a review to IEEE Transaction on Communication).  Personal use of this material is permitted.  Permission from IEEE must be obtained for all other uses, in any current or future media, including reprinting/republishing this material for advertising or promotional purposes, creating new collective works, for resale or redistribution to servers or lists, or reuse of any copyrighted component of this work in other works.”}

\clearpage


\section{Introduction}
Artificial Neural Networks (ANNs) is a computational model that endures input information and produces output data subject to their predefined activation functions. In practice, the ANN model is trained either practicing the labeled or unlabeled data. In the recent years, Deep Learning (DL), which is a sub-branch of Machine Learning (ML) [1], has attracted numerous researchers to investigate it's application in communication. The ANN is the essential structure square of DL in such applications. Researchers incorporated DL in communication system during various processes, such as modulation [3], channel estimation [4-6], signal detection [7, 8], modulation recognition [9, 10], and channel decoding [11].

Convolutional Neural Networks (CNNs) [2] are a type of DL model that showed enormous accomplishment in the fields of the object, face, handwriting, speech, emotion recognition, and language processing [1]. However, CNN requires a large amount of labeled data during the training phase; therefore, in the absence of sufficient training data, the accuracy is degraded. To improve the accuracy, Autoencoder (AE) is employed demonstrating the empowered accuracy consuming limited data during the training phase [12]. Also, an AE architecture performs encoding and modulation at the transmitter, whereas decoding and demodulation at the receiver. An AE [12] is a type of ANN that has gained prominence as it is trained with unlabeled data and is employed in learning efficient data encoding. The encoder intends to learn a dataset representation through the mapping to the low dimensional space. The decoder part attempts to regenerate the original data by ignoring noise such that the output data is obtained nearest to input. The encoder and decoder of an AE are defined as transitions $\phi$ and $\psi$  as given in Eq.(1) [17].

\begin{figure}[H]
\centering
\captionsetup{justification=centering,margin=2cm}
\centerline{\includegraphics[scale=1.4]{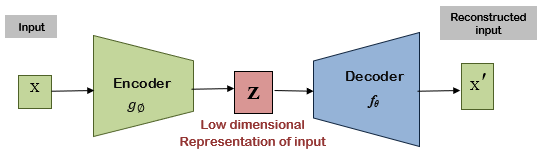}}
\caption{Architecture of an AE incorporated communication system.}
\label{fig}
\end{figure}
\begin{equation}
\begin{aligned}
&\phi: \mathcal{X} \rightarrow \mathcal{F}\\
&\psi: \mathcal{F} \rightarrow \mathcal{Y}
\end{aligned}
\end{equation}

Where X represents the furnished input, and Y represents the predicted output. The transition $\phi$ is chosen, such that the mapping from X results in lower-dimensional Z in feature space F.

\section{Previous work} 

In practice, ML is incorporated in numerous applications in communication systems, such as modulation recognition, spectrum sensing, and prediction [19-20]. The authors in [23-25] have exercised Recurrent Neural Network (RNNs) at the receiver to decode convolutional, Polar, and Turbo codes, which are channel-coded information bits. Besides, the authors in [24-25] demonstrated that RNNs effectively extricate the information from an extended sequence in the presence of higher channel noise. Additionally, it learns the underlying architecture of the human-designed programs with the limited training data; however, the consideration of a transmitter processing, or a complete end-to-end system design applying RNNs remains unexplored. On the other hand, with the evolution of ML, a single neural network can represent an entire communication system. One such system is an AE, which is used in [16] and proved that the transmitter and receiver are optimized simultaneously employing an AE's framework. The AE's system [16] has improved significantly in [21] with the introduction of radio transformer networks [20], which helps in actualizing communication knowledge into DL models. The authors in [22] exhibited that the complex neural networks could be actualized on the hardware. Following this, in recent years, AE-based communication systems were established employing convolutional layers as the essential elements, relaxing the restriction on input bit sequence length [14-15]. However, the performance in[14]isaffected by an error floor contemplating the high Signal to Noise Ratio (SNR) regime because the network was implemented employing ML. Considering this drawback, the authors in [13] proposed a generalized system examining the various parameters, such as throughput, block length, and channel occupancy.

In general, AE employees Restricted Boltzmann Machine (RBM) or Gaussian-RBM (G-RBM) activation sigmoid function between the input and the hidden layer, causing the shortcoming of the vanishing gradient. To circumvent this, we employ the Variational Autoencoder (VAE) [18] during the advanced training phase in the proposed work. Furthermore, we introduced the tunable parameter $\beta$ in the original VAE, which is also recognized as $\beta-$ VAE, exhibiting enhanced disentangled latent space representation.

The remainder of the paper is composed as follows. Section III presents the system model, Section IV illustrates VAE oriented communication system architecture, while Section V explicates simulation and results, and finally, Section VI concludes the paper. 

\section{System Model}
The proposed architecture of the communication system is inspired by Auto Encoding Variational Bayes (AEVBs) [19]. To describe the working principle of AEVBs, assume a data vector $\mathbf{x}$, which is encoded into a vector $\mathbf{h}$ using an encoding distribution $\mathbf{e}_{\phi}$ parametrized by $\phi$. The encoded data vector $\mathbf{h}$ is transmitted through channel represented by the function $\mathbf{c(\cdot)}$, and the received data vector is $\mathbf{c(h)}$. Let, $\mathbf{d}_{\theta}$ be a distribution, parametrized by $\mathbf{\theta}$ that decodes the received data vector $\mathbf{c(h)}$ into the original data vector $\mathbf{x}$.

AEVBs [19], unlike the AEs, employs the Kullback-Leibler (KL)-divergence between the predicted probability distribution function and the actual posterior distribution in the latent space. In contrast, traditional AEs attempt to obtain the proper mapping functions within the encoder (between input and latent space) as well as within the decoder (between the latent space and output). Thus, if a system is tested in the presence of recent input or input that is profoundly affected by the noise, AE's accurate mapping is compromised. Therefore, employing AEVB, we generate a large dataset by combining the noise in the latent space, which is similar to input data augmentation (combining noise to images increasing the number of examples in the input dataset). However, AEVB introduces an additional loss element as a consequence of the variational approach in latent representation learning.

For the various choice of the inference model $e_{\phi}(\mathbf{h} | \mathbf{x})$ and variational parameter ${\phi}$, we acquire:
\begin{equation}
\begin{aligned}
\log d_{\boldsymbol{\theta}}(\mathbf{x}) &=\mathbb{E}{e_{\phi}(\mathbf{h} | \mathbf{x})}\left[\log d_{\boldsymbol{\theta}}(\mathbf{x})\right] \\
&=\mathbb{E}{e_{\phi}(\mathbf{h} | \mathbf{x})}\left[\log \left[\frac{d_{\boldsymbol{\theta}}(\mathbf{x}, \mathbf{c(h)})}{d_{\boldsymbol{\theta}}(\mathbf{c(h)} | \mathbf{x})}\right]\right] \\
&=\mathbb{E}{e_{\phi}(\mathbf{h} | \mathbf{x})}\left[\log \left[\frac{d_{\boldsymbol{\theta}}(\mathbf{x}, \mathbf{c(h)})}{e_{\boldsymbol{\phi}}(\mathbf{h} | \mathbf{x})} \frac{e_{\boldsymbol{\phi}}(\mathbf{h} | \mathbf{x})}{d_{\boldsymbol{\theta}}(\mathbf{c(h)} | \mathbf{x})}\right]\right]   \\ &={\mathbb{E}{e_{\phi}(\mathbf{h} | \mathbf{x})}\left[\log \left[\frac{d_{\boldsymbol{\theta}}(\mathbf{x}, \mathbf{c(h)})}{e_{\boldsymbol{\phi}}(\mathbf{h} | \mathbf{x})}\right]\right]}
\\ &+{\mathbb{E}{e_{\phi}(\mathbf{h} | \mathbf{x})}\left[\log \left[\frac{e_{\phi}(\mathbf{h} | \mathbf{x})}{d_{\boldsymbol{\theta}}(\mathbf{c(h)} | \mathbf{x})}\right]\right]}
\\ & {=\mathcal{L({\boldsymbol{\theta}}, \phi}; \mathbf{x}}){+D_{K L}\left(e_{\phi}(\mathbf{h} | \mathbf{x})||d_{\boldsymbol{\theta}}(\mathbf{c(h)} | \mathbf{x})\right)}
\end{aligned}
\end{equation}

In Eq.(2) [19], the first term represents the variational lower bound, which is also called the Evidence Lower Bound (ELBO), and the second term represents the KL-divergence between $e_{\phi}(\mathbf{h} | \mathbf{x})$ and $d_{\theta}(\mathbf{c(h)} | \mathbf{x})$, which is expected to be positive as indecated in Eq.(3).
\begin{equation}
D_{K L}\left(e_{\phi}(\mathbf{h} | \mathbf{x}) \| d_{\theta}(\mathbf{c(h)} | \mathbf{x})\right) \geq 0
\end{equation}

On the other hand, Eq.(3) indicates if $e_{\phi}(\mathbf{h} | \mathbf{x})$ equals to the true distribution $d(c(h)|x)$, KL-divergence remains zero. 
\begin{equation}
\mathcal{L}({\boldsymbol{\theta}, \boldsymbol{\phi}}; \mathbf{x})=\mathbb{E}{e_{\phi}(\mathbf{h} | \mathbf{x})}\left[\log d_{\boldsymbol{\theta}}(\mathbf{x}, \mathbf{c(h)})-\log e_{\phi}(\mathbf{h} | \mathbf{x})\right]
\end{equation}

Eq.(4) presents the lower bound. As a consequence of the non-negativity nature of the KL-divergence, the ELBO on the log-likelihood of the data is presented in Eq.(5).
\begin{equation}
\begin{aligned}
\mathcal{L}({\boldsymbol{\theta}, \boldsymbol{\phi}}; \mathbf{x}) &=\log d_{\boldsymbol{\theta}}(\mathbf{x})-D_{K L}\left(e_{\phi}(\mathbf{h} | \mathbf{x}) \| d_{\boldsymbol{\theta}}(\mathbf{c(h)} | \mathbf{x})\right) \\
& \leq \log d_{\boldsymbol{\theta}}(\mathbf{x})
\end{aligned}
\end{equation}

Interestingly, the KL-divergence ascertains two distances:

1. The KL-divergence of the approximate posterior
from the correct posterior;

2. The difference between the ELBO $d_{\theta, \phi}( \mathbf{x})$, and the marginal likelihood $log(d_{\theta}( \mathbf{x}))$; which is also known as the tightness of the bound. The better $e_{\phi}(\mathbf{h} | \mathbf{x})$ approximates the correct (posterior) distribution $d_{\theta}(\mathbf{c(h)} | \mathbf{x})$, in terms of the KL-divergence, with the reduced difference.

In this case, assuming data is generated employing $ d_{\theta}( \mathbf{x}|\mathbf{c(h)})$ and that the encoder is also trained through learning an approximation $e_{\phi}(\mathbf{h} | \mathbf{x})$ of the posterior distribution $ d_{\theta}( \mathbf{x}|\mathbf{c(h)})$, where $\phi$ and $\theta$ indicate the encoder and decoder parameters, respectively. Therefore,  the conditional or posterior distribution is rewritten in Eq.(6).
\begin{equation}
d_{\boldsymbol{\theta}}(c(h) | x)=\frac{d_{\boldsymbol{\theta}}(c(h), x)}{d_{\boldsymbol{\theta}}(x)}=\frac{d_{\boldsymbol{\theta}}(x | c(h)) d_{\boldsymbol{\theta}}(c(h))}{d_{\boldsymbol{\theta}}(x)}
\end{equation}

Eq.(6) denominates the marginal distribution of observations and is calculated through marginalizing the latent space. Therefore, Eq. (7) depicts that AEVBs minimize the gap between $e_{\phi}(\mathbf{h} | \mathbf{x})$ and the posterior distribution $d_{\theta}(\mathbf{c(h)} | \mathbf{x})$ such the received data is relatively consistent with input data. Furthermore, Eq.(8) presents the objective of the AEVBs, where $D_{KL}$ represents the KL-divergence. 
\begin{equation}
\text {Minimize: } D_{K L}\left[e_{\phi}(h | x) \| d_{\theta}(c(h) | x)\right]
\end{equation}
\begin{equation}
\begin{aligned}
\mathcal{L}(\phi, \theta, \mathbf{x}) &= 
\mathbf{D}_{\mathrm{KL}}\left(\mathbf{e}_{\phi}(\mathbf{h} | \mathbf{x}) \| \mathbf{d}_{\theta}(\mathbf{c(h)})\right)\\
&-\mathbb{E}{e_{\phi}(\mathbf{h} | \mathbf{x})}\left(\log \mathbf{d}_{\theta}(\mathbf{x} | \mathbf{c(h)})\right)
\end{aligned}
\end{equation}

\textbf{Theorem 1:} Let, the latent vector $\widetilde{\mathbf{h}}$, sampled from the encoder distribution $\mathbf{e}_{\phi}(\mathbf{h} | \mathbf{x})$, is reparametrized employing a differential transformation $\mathbf{r}_{\phi}(\mathbf{\epsilon}, \mathbf{x})$, where $\mathbf{\epsilon}$ is sampled from a predefined distribution $\mathbf{p}(\mathbf{\epsilon})$: $\widetilde{\mathbf{h}} = \mathbf{r}_{\phi}(\mathbf{\epsilon}, \mathbf{x})$ where $\mathbf{\epsilon} \sim \mathbf{p}(\mathbf{\epsilon})$. The expectation in the loss $\mathcal{L}(\phi, \theta, \mathbf{x})$
is approximated using Monte-Carlo estimate as shown in Eq.(9).
\begin{equation}
\begin{aligned}
\mathbb{E}_{\mathbf{e}_{\phi}\left(\mathbf{h} | \mathbf{x}\right)}[f(\mathbf{c(h)})] &= \mathbb{E}_{\mathbf{p}(\epsilon)}\left[f\left(\mathbf{c}\left(\mathbf{r}_{\phi}\left(\epsilon, \mathbf{x}\right)\right)\right)\right] \\
&\simeq \frac{1}{N} \sum_{n=1}^{N} f\left(\mathbf{c}\left(\mathbf{r}_{\phi}\left(\epsilon^{(n)}, \mathbf{x}\right)\right)\right) \\
\quad \text { where } \quad \epsilon^{(n)} \sim p(\epsilon)
\end{aligned}
\end{equation}

Considering the \textbf{Theorem 1}, the loss function $\mathcal{L}(\phi, \theta, \mathbf{x})$ is further approximated and presented in Eq.(10).
\begin{equation}
\begin{aligned}
\mathcal{L}(\phi, \theta, \mathbf{x}) & \simeq 
\mathbf{D}_{\mathrm{KL}}\left(\mathbf{e}_{\phi}(\mathbf{h} | \mathbf{x}) \| \mathbf{d}_{\theta}(\mathbf{c(h)})\right)\\
&-\frac{1}{N} \sum_{n=1}^{N} \log \mathbf{d}_{\theta}\left(\mathbf{x} | \mathbf{c}\left(\mathbf{h}^{(n)}\right)\right)
\end{aligned}
\end{equation}

$\text { where } \quad \mathbf{h}^{(n)} = \mathbf{r}_{\phi}(\mathbf{\epsilon}, \mathbf{x}) \quad \text {and} \quad \epsilon^{(n)} \sim p(\epsilon)$

\subsection{Variational Autoencoder (VAE)}
In practice, to optimize the Eq.(10), we assume that the prior $d(h)$ distribution is typically positioned as the isotropic unit of Gaussian $\mathcal{N}(0,1)$, and posterior distributions are parametrized as Gaussians with a diagonal covariance matrix. This allows the use of the “reparametrisation trick” estimating the lower bound gradients conserning the parameters $\phi$, where each of the random variables $h_{i}\sim$ $e_{\phi}\left(h_{i} | \mathbf{x}\right)=\mathcal{N}\left(\mu_{i}, \sigma_{i}\right)$ is parametrised as the differentiable transformation of the noise variable $\epsilon \sim$ $\mathcal{N}(0,1)$ as shown in Eq.(11).
\begin{equation}
    \mathbf{h}_{i}=\mu_{i}+\sigma_{i} \epsilon
\end{equation}

\textbf{Theorem 2:} If the prior distribution $d(h)$ is isotropic unit of Gaussian $\mathcal{N}(0,1)$ and for the encoder distribution $e_{\phi}\left(h_{i} | \mathbf{x}\right)=\mathcal{N}\left(\mu_{i}, \sigma_{i}\right)$
the KL-Divergence of the loss $\mathcal{L}(\phi, \theta, \mathbf{x})$ is computed in Eq.(12) and proved in Appendix B of [19].
\begin{equation}
\begin{aligned}
    \mathbf{D}_{\mathrm{KL}}\left(\mathbf{e}_{\phi}(\mathbf{h} | \mathbf{x}) \| \mathbf{d}_{\theta}(\mathbf{c(h)})\right) = -\frac{1}{2} \sum_{j=1}^{J}\left(1+\log \left(\sigma_{j}^{2}\right)-\mu_{j}^{2}-\sigma_{j}^{2}\right)
\end{aligned}
\end{equation}

This assumption of both $d_{\theta}(\mathbf{h})$ (the prior) and $e_{\phi}(\mathbf{h} | \mathbf{x})$ are Gaussian leads to a model named VAE, which and differentiate the KL-divergence employing Eq.(10) and \textbf{Theorem 2}. Furthermore, the resulting loss, considering the Eq.(10), is derived in Eq.(13).
\begin{equation}
\begin{aligned}
\mathcal{L}(\boldsymbol{\theta}, \boldsymbol{\phi} ; \boldsymbol{x}) & \simeq  \frac{1}{2} \sum_{j=1}^{J}\left(1+\log \left(\sigma_{j}^{2}\right)-\mu_{j}^{2}-\sigma_{j}^{2}\right) \\
&+\frac{1}{N} \sum_{i=1}^{N} \log \mathbf{p}_{\theta}\left(x | c(h^{(n))}\right)
\end{aligned}
\end{equation}

where $\quad \mathbf{h}^{(n)}=\boldsymbol{\mu}+\boldsymbol{\sigma} \odot \boldsymbol{\epsilon}^{(n)}$ and $\quad \boldsymbol{\epsilon}^{(n)} \sim \mathcal{N}(0, \mathbf{I})$

\begin{figure}[htbp]
\centering
\captionsetup{justification=centering,margin=2cm}
\centerline{\includegraphics[scale =0.9]{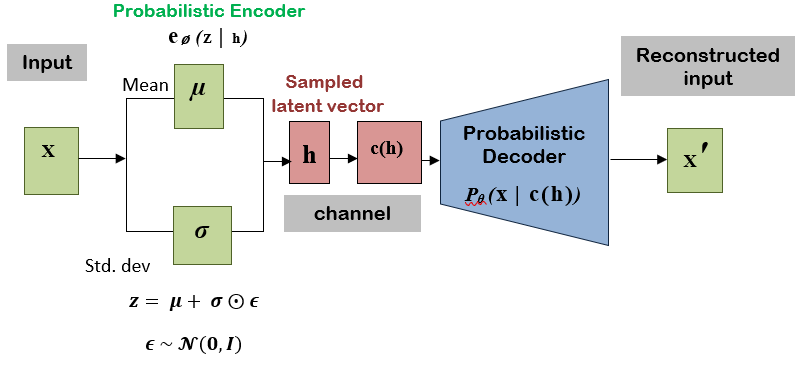}}
\caption{The architecture of a VAE based communication system.}
\label{fig}
\end{figure}

\subsection{$\beta$-VAE}
To overcome the reconstruction-disentanglement trade-off, achieve higher robust disentangling, and greater reconstruction accuracy, we introduce a hyperparameter $\beta$ to the proposed VAE architecture, which is recognized as $\beta$-VAE. Therefore the Eq.(8) is further modified and presented in Eq.(14).
\begin{equation}
\begin{aligned}
     \mathcal{L}(\boldsymbol{\theta}, \boldsymbol{\phi} ; \mathbf{x}, \mathbf{h}, \beta) = 
     &\beta \mathbf{D}_{K L}\left(\mathbf{e}_{\phi}(\mathbf{h} | \mathbf{x}) \| \mathbf{d}(\mathbf{c(h)})\right)\\
     &-\mathbb{E}{e{\phi}(\mathbf{h} | \mathbf{x})}\left[\log \mathbf{d}_{\theta}(\mathbf{x} | \mathbf{c(h)})\right]
\end{aligned}
\end {equation}

The value of $\beta$ should be chosen in such that it results in more disentangled latent representations $\mathbf{h}$. Nevertheless, the earlier study suggested that increasing the value of $\beta$ to accomplish highly disentangled representations inferred in additional capacity constraints of the latent bottleneck $\mathbf{h}$ and extra pressures for it to be factorized while still being sufficient to reconstruct the data $\mathbf{x}$. Therefore the value of $\beta$ is efficiently chosen to maintain the trade-off among the fidelity of $\beta$-VAE reconstructions and the disentangled nature of its latent space $\mathbf{c(h)}$. The value of $\beta$ is chosen $10^{-4}$ across all experiments.

\begin{figure}[htbp]
\centerline{\includegraphics[width=7 cm ,height =15cm]{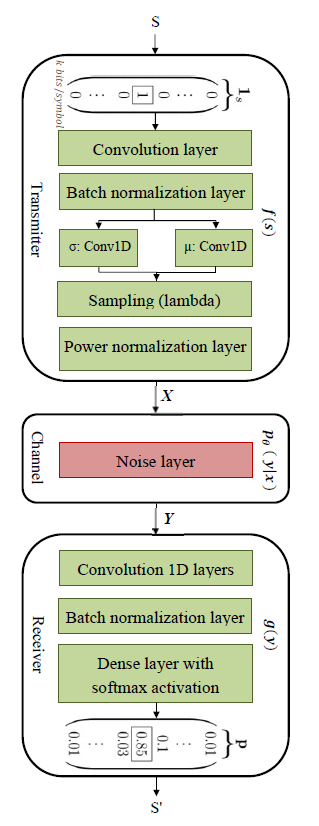}}
\caption{The block diagram of the proposed AEVB oriented communication system. The symbol S is coded as a one-hot input vector, whereas S' is the output vector, which is picked up considering the probability distribution across all the possible messages.}
    \label{fig}
\end{figure}
\section{VAE Oriented Communication System}
\subsection{Network Architecture}

A communication system presented in Fig. 2 and Fig. 3 consists of a transmitter, channel, and receiver.  Transmitter acquires the input symbol S  comprised of k information bits acquired from the client source. Fig. 2 and Fig. 3 depict that the channel propagates the signal, and the receiver receives the transmitted signal from the channel. Moreover, the transmitter's primary function is to transform a symbol S to a signal X such that X occupies n time slots in the channel. The generated signal from the transmitter is subjected to power constraints in the channel; as a result, there is a possibility of obtaining a noisy and distorted signal at the receiver side. Hence, the foremost responsibility of the receiver is to estimate the symbol as close as possible to the original symbol S. Therefore, considering the generalized VAE architecture, as depicted in Fig. 2, we proposed a CNN-based VAE communication system (including an adjustable hyperparameter $\beta$, which is present in the sampling layer), as presented in Fig. 3. The system code rate is R= k/n (bits/channel-used).

The proposed system consists of five 1-dimensional convolutional (Conv1D) layers, three in the transmitter, and two in the receiver. The transmitter consists of a batch normalization layer, sampling, and power normalization, which ensures that power constraints on the generated signal X are satisfied. In addition to two Conv1D layers, the receiver also contains a batch normalization layer and a softmax layer to classify the received data as 0 or 1. Although increasing the number of convolution layers ensures the increase in the neural network's classification power, it also leads to gradient exploding and increased parameters to train, which increases the computational complexity. In the current work, we have achieved the improved BLER performance employing three Conv1D layers in the transmitter, reducing the number of training parameters to 12,824, which is substantially lesser compared to 145,434 in [13], Furthermore, employing their Conv1D layers at the transmitter avoids the gradient exploding and lessens the complexity.

The transmitter converts the source input symbol into a one-hot vector before further processing, enhancing the BLER performance. Moreover, the Conv1D layer ensures that the block of symbols is processed instead of bits, conceding us to study BLER.  In the proposed work, it can process a total of k*L bits simultaneously where k represents a number of bits per symbol, and L represents the number of symbols per block length. Therefore BLER, in this case, is defined as the number of symbols received correctly out of L symbols. The detailed parameters employed in the corresponding layer of the proposed CNN-based AE communication system are presented in Fig.4.

As per the channel coding theorem, as presented in Fig. 4, the Conv1D layer transforms the one-hot input symbol sequence into a distinct signal representation X occupying the n channels. Employing the DL method, each Conv1D layer contains 256 filters, enabling the mapping of one hot symbol sequence to 256-dimensional space. Besides, the sampling layer is employed to learn a data generating distribution, enabling us to exert the random samples from the latent space. Hence, the transmitter is learning an approximation to the posterior distribution ($ p( \mathbf{X})$). The power normalization layer in the transmitter is required to compress the 256-dimensional representation furnished by the convolution layer to a 2n-dimensional space signal of ${X}$, representing the signal constellation diagram of the input symbol sequence. The parameters employed in the power normalization layer are programmable to accomplish the mapping to 4n dimensional space as the higher dimensional latent space produces sounder latent representation and enhanced pattern recognition respecting the input information. On the other hand, increasing the latent space dimensions further increases the computational complexity and transmission cost per bit.

The conditional probability density function $ p( \mathbf{y}|\mathbf{x})$ () represents the channel layer in the presence of noises. The current study implements an AWGN channel with a fixed variance $\sigma_{1}=\left(2 R E_{b} / N_{0}\right)^{-1}$ and flat Rayleigh fading channel, performing convolution of the transformed signal X with channel impulse responses.  The receiver classifies the received signal ${Y}$, where ${n}$ represents the noise. In the proposed architecture, the receiver consists of two Conv1D layers. The first layer decompresses the received signal ${Y}$ back to 256-dimensional space to reconstruct the transmitted information. The second layer employs the signal representation data to classify the received symbol correctly. Finally, the signal is mapped into a one-hot vector of original length k performing the soft decision employing the soft-max activation.

\subsection{System Parameters}
The proposed architecture is trained and tested through a random dataset generated using a uniform distribution. Furthermore, the generated dataset comprises of training and testing sets, consisting of 12800 and 64000 data messages setting a batch size of 64, respectively.  Each information message consists of L symbols, and each symbol consists of k information bits. The Conv1D layers share the property of weight factor, and hence the CNN-VAE system is trained and tested on the arbitrary block length L. However, for simplicity, we set the block length L to 100 and the adjustable parameter $\beta$ to $10^{-4}$. For instance, Eb/No (SNR) is considered in the range of 5 to 15 dB during the training and testing phase. For comparison, the simulated BLERs are presented considering 2n and 4n dimensions, inferencing that the latent space representation in the 4n dimension is ameliorated than the 2n. Moreover, Fig. 4 presents the complete block diagram revealing the key parameters,  such as kernel and stride size in the convolution layers. Whereas "a," "b," and "c" represent the batch, kernel size, and stride sizes, respectively. Furthermore, a loss function is defined as the binary cross-entropy because the sequences (input and output) are converted to one-hot vectors. An Adam optimizer is employed in training the CNN-VAE system, and the learning rate is fixed to 0.01. The training is performed over 150 epochs such that the system rapidly converges employing the batch normalization layer. Since the proposed system is memoryless and learns only the underlying structure in the data, it subsequently decodes unseen codewords, which justifies training on 12800 and testing on 68000 data messages, respectively.
 
 \begin{figure}[H]
\centerline{\includegraphics[width=16 cm,height =16 cm]{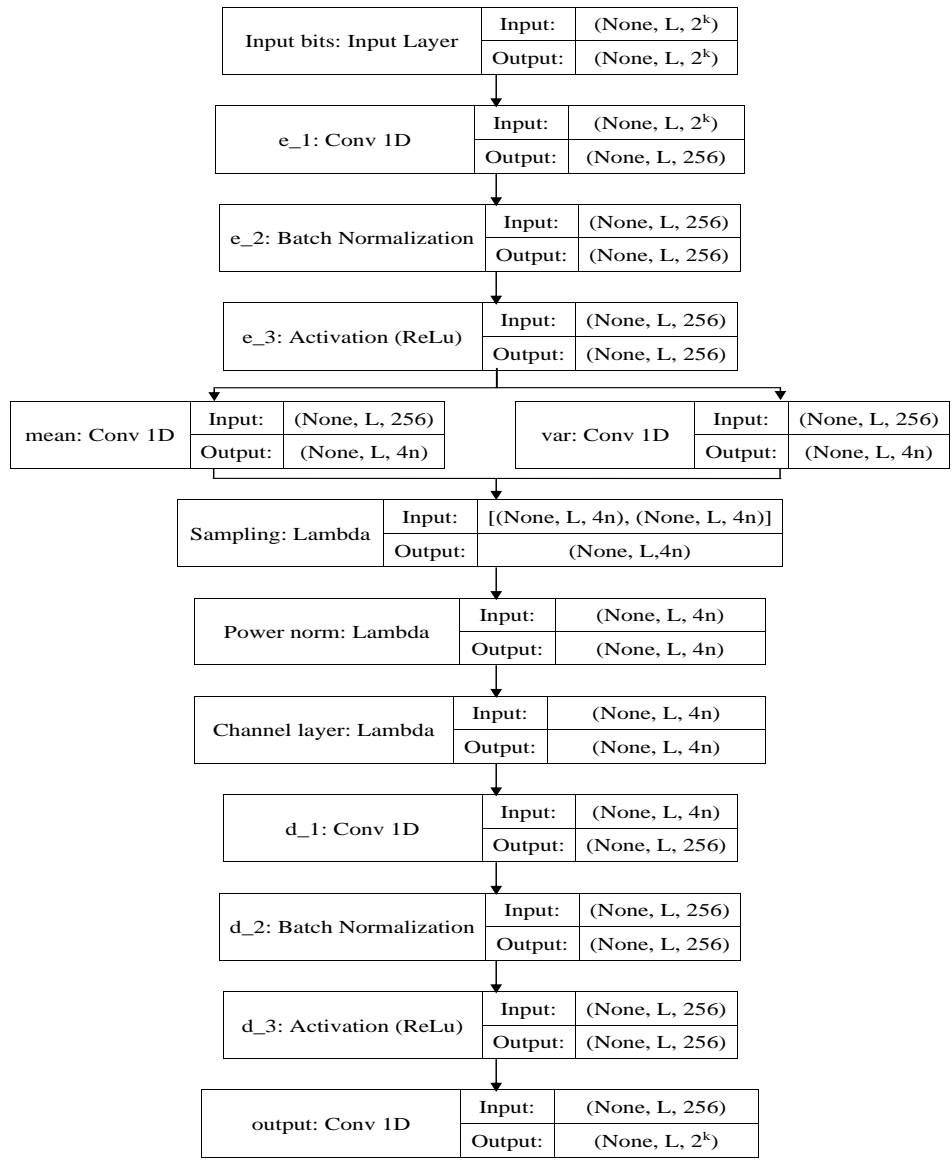}}
\caption{Block diagram of the proposed communication system with the respective layer settings.}
\label{fig}
\end{figure}

\section{Simulation Results}

This section presents the comprehensive simulation and results for the proposed AEVBs based communication system considering the AWGN and Rayleigh fading channels.
\subsection{AWGN Channel}
Initially, we trained the system at a single Eb/No while testing over the complete Eb/No range.  As shown in Fig.  5, the system is trained at 6 and 9dB of Eb/No, with the code rates of R=2 and 4 bits/channel-used and 2n and 4n latent space representation dimensions. Fig. 5 also demonstrates that the proposed system achieves the improved BLER performance when trained at 6dB of Eb/No. Additionally, the system is trained considering numerous samples enabling the receiver to learn the underlying structure efficiency in the presence of higher noise. 

\begin{figure}[htbp]
\centerline{\includegraphics[scale=1.2]{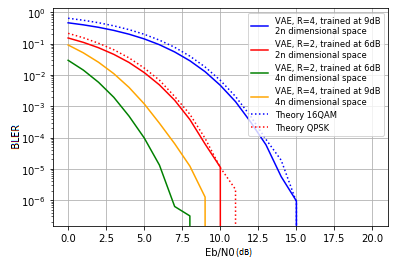}}
\caption{Performance analysis of the proposed system (Fig. 3) in terms of BLER, by considering two different code rates R=2 and 4 (bits/channel-used), comparing with the 16QAM and QPSK under AWGN channels.}
\label{fig}
\end{figure}

Fig. 6 and 7 exhibit the simulation results for the system, which is trained at two different code rates k=4, k=8 (information bits/channel), two AWGN channel employed n=1, n=2, under 2n, and 4n latent space dimensions. It is evident that an increased number of time slots $(n>1)$ at the channel for each symbol transmission, enables convolution layers at the transmitter to utilize the additional time-domain resources, increasing the system performance. Moreover, Fig. 6 and 7revealthat 4n dimension representation of latent space accomplishes improved BLER performance compared to 2n dimension, which justifies that higher-dimensional latent space representation has enhanced performance.
\begin{figure}[htbp]
\centerline{\includegraphics[scale=1.2]{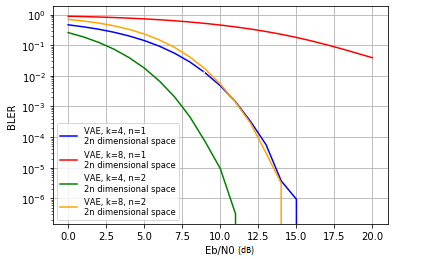}}
\caption{Performance analysis of the proposed system (Fig. 3) in terms of BLER, by considering two different channel usage (n=1, n=2) and two different data lengths (k=4, k=8) with latent space represented in 2n dimension under AWGN channel.}
\label{fig}
\end{figure}

\begin{figure}[htbp]
\centerline{\includegraphics[scale=1.2]{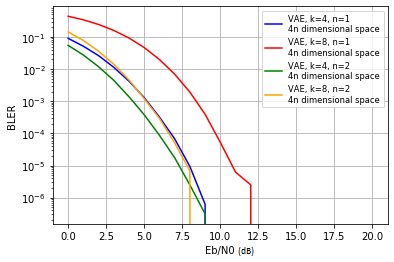}}
\caption{Performance analysis of the proposed system (Fig. 3) in terms of BLER, by considering two different channel usage (n=1, n=2) and two different data lengths (k=4, k=8) with latent space represented in 4n dimension under AWGN channels.}
\label{fig}
\end{figure}
Fig.  8 demonstrates the performance comparison of the proposed and existing systems. The proposed system is trained considering two different code rates and two different 2n and 4n latent space dimensions. Fig. 8 illustrates that the performance of the proposed system outclasses the traditional systems. 
\begin{figure}[htbp]
\centerline{\includegraphics[scale=1.2]{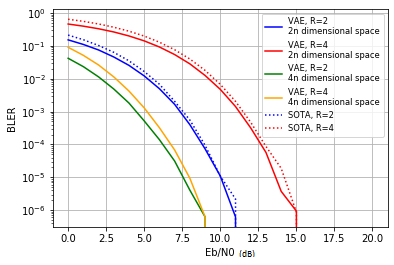}}
\caption{Performance analysis of the proposed system (Fig. 3) in terms of BLER, by considering two different code rates R=2 and 4 (bits/channel-used) and two different latent space dimensions when compared with the state of the art (SOTA) under AWGN.}
\label{fig}
\end{figure}

\begin{figure}[H]
\centerline{\includegraphics[scale=1.2]{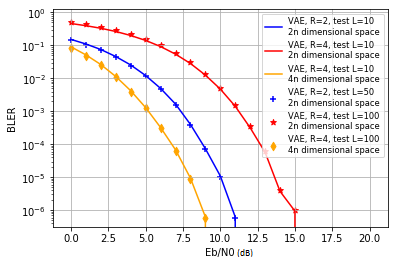}}
\caption{Performance analysis of the proposed system (Fig. 3) in terms of BLER, by considering two different code rates R=2 and 4 (bits/channel-used) trained at block length L=100 and tested at L=10, L=50, and L=100 under AWGN.}
\label{fig}
\end{figure}  
As we cannot estimate the block length in future versions and to provide the generalization capability of the proposed architecture in terms of handling different block lengths , we have trained the system presented in Fig.3 employing a block length (L=10) and further obtained the BLER for the several other block lengths, such as L=50 and100 under AWGN adopting the aforementioned parameters. The results in Fig.9 explicitly demonstrated the same thing by achieving identical performance.
  
\subsection{Rayleigh Fading Channel}
Fig. 10 presents the performance analysis of the proposed system (Fig. 3) in terms of BLER, by considering two different code rates R=2 and 4 (bits/channel-used), when comparing with the corresponding 16QAM and QPSK under the Rayleigh fading channels.
\begin{figure}[H]
\centerline{\includegraphics[scale=1.1]{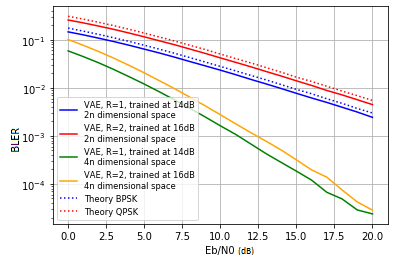}}
\caption{Performance analysis of the proposed system (Fig. 3) in terms of BLER, by considering two different code rates R=2 and 4 (bits/channel-used), when comparing with the corresponding 16QAM and QPSK under the Rayleigh fading channels.}
\label{fig}
\end{figure}

All the explanations for the results of Fig.  6 and 7 work fine with Fig.11 and 12 except that they only vary in the channel noise, former being AWGN and later being Rayleigh fading channel.
\begin{figure}[htbp]
\centerline{\includegraphics[scale=1.2]{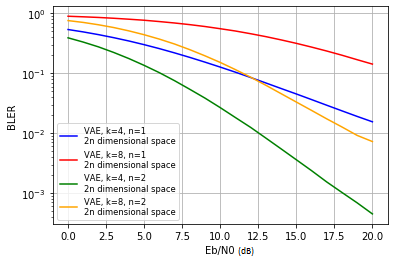}}
\caption{Performance analysis of the proposed system (Fig. 3) in terms of BLER, by considering two different channel usage (n=1, n=2) and two different data lengths (k=1, k=2) with latent space represented in 2n dimension under Rayleigh fading channels.}
\label{fig}
\end{figure}

\begin{figure}[htbp]
\centerline{\includegraphics[scale=1.2]{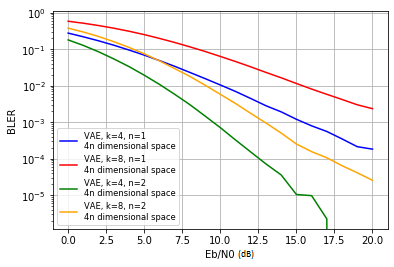}}
\caption{Performance analysis of the proposed system (Fig. 3) in terms of BLER, by considering two different channel usage (n=1, n=2) and two different data lengths (k=1, k=2) with latent space represented in 4n dimension under Rayleigh fading channels.}
\label{fig}
\end{figure}

Fig.13 confers enhanced performance for the proposed system compared to the traditional architectures in all the four combinations, i.e. when trained at two different code rates and two different dimensional spaces by considering Rayleigh fading channels.

\begin{figure}[htbp]
\centerline{\includegraphics[scale=1.2]{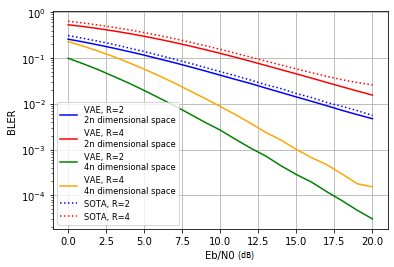}}
\caption{Performance analysis of the proposed system (Fig. 3) in terms of BLER, by considering two different code rates R=2, 4 (bits/channel-used) and two different latent space dimensions when compared with the SOTA under Rayleigh fading channels.}
\label{fig}
\end{figure}
Fig. 14 proves that the proposed system achieves identical performance in terms of BLER over different block lengths even under Rayleigh fading channels, with the trained communication system (Fig. 3) employing a block length (L=10) and obtained the BLER considering the several block length, such as L=10, 50, and 100 employing the parameters, as mentioned earlier. The results in Fig.14 also demonstrates that even under Rayleigh fading channels, the system performed excellently.
\begin{figure}[htbp]
\centerline{\includegraphics[scale=1.2]{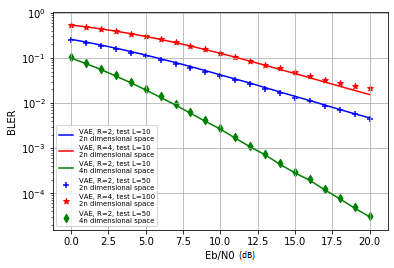}}
\caption{Performance analysis of the proposed system (Fig. 3) in terms of BLER, by considering two different code rates R=2 and 4 (bits/channel-used) trained at block length L=100 and tested at L=10, L=50, and L=100 under Rayleigh fading channels.}
\label{fig}
\end{figure}

\subsection{Training Convergence}
The proposed CNN-VAE system demonstrates rapid convergence. Fig. 15 demonstrates the training and validation loss concerning the training epochs under AWGN, and it is observed that the system was insensitive to initialization parameters but converged for almost 20 epochs, and it is almost found to be constant till 150 epochs.
\begin{figure}[H]
\centerline{\includegraphics[scale=0.8]{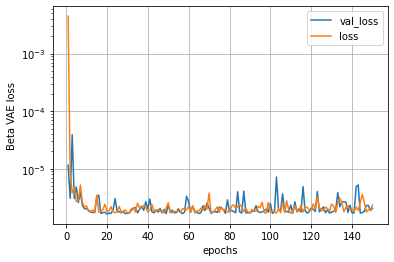}}
\caption{Training and validation losses of the proposed system under AWGN added channel with data length k=4 and using 150 epochs to train.}
\label{fig}
\end{figure}

Training and validation losses as a function of training epochs under the Rayleigh fading channel is shown in Fig. 16, and it is observed that the system was insensitive to initialization parameters but converged rapidly in less than 20 epochs, and remains constant till 150 epochs.
\begin{figure}[H]
\centerline{\includegraphics[scale=0.8]{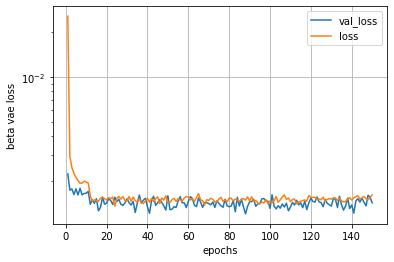}}
\caption{Training and validation losses for the proposed system under the Rayleigh fading channels, with data length k=4 and using 150 training epochs.}
\label{fig}
\end{figure}

\section{conclusion}

The present paper has proposed the novel communication system adopting the CNN based AE architecture. Employed convolution layers at the transmitter and receiver have performed the necessary coding and decoding. An adjustable hyperparameter $ \beta $ has been incorporated to increase the reconstruction accuracy and disentangle the latent space. The number of trainable parameters has found to be very less, which ensures that the proposed architecture reduces the complexity. The performance of the CNN-VAE (proposed) system is analyzed with randomly generated data under AWGN and Rayleigh channels with the variation in the block lengths, code rates, channel-used n, and SNR (Eb/No). Furthermore, it observed that the 4n dimensional representation of latent space showed improved BLER performance compared to 2n dimensional representation. Finally, the proposed architecture can converge rapidly in the limited epochs. Therefore, the proposed system can communicate intelligently under AWGN and Rayleigh fading channels and accomplishes enhanced performance compared to the traditional systems.


\begin{thebibliography}{00}
\bibitem{b1} Y. LeCun et al., “Deep learning,” Nature, vol. 521, pp. 436-444, May 2015.
\bibitem{b2} P. S. Padwal et al., “Survey on Convolutional Neural Networks,” vol. 7, no. 6, Jun. 2018.
\bibitem{b3} A. Felix et al., “OFDM-Autoencoder for End-to-End Learning of Communications Systems,” IEEE 19th International Workshop on Signal Processing Advances in Wireless Communications (SPAWC), Kalamata, Greece, Jun. 2018, pp. 1-5.
\bibitem{b4} H. He et al., “Deep Learning-Based Channel Estimation for Beamspace mmWave Massive MIMO Systems,” IEEE Wireless Communications Letters, vol. 7, no. 5, pp. 852-855, Oct. 2018.
\bibitem{b5} D. Neumann, T. Wiese, and W. Utschick, “Learning the MMSE Channel Estimator,” IEEE Transactions on Signal Processing, vol. 66, no. 11, pp. 2905-2917, Jun., 2018.
\bibitem{b6} H. Ye et al., “Deep Learning-Based End-to-End Wireless Communication Systems With Conditional GANs as Unknown Channels,” IEEE Transactions on Wireless Communications, vol. 19, no. 5, pp. 3133-3143, May 2020.
\bibitem{b7} N. Samuel, T. Diskin, and A. Wiesel, “Deep MIMO detection,” IEEE 18th International Workshop on Signal Processing Advances in Wireless Communications (SPAWC), Sapporo, Japan, Dec. 2017, pp. 1-5.
\bibitem{b8} H. Ye et al., “Power of Deep Learning for Channel Estimation and Signal Detection in OFDM Systems,” IEEE Wireless Communications Letters, vol. 7, no. 1, pp. 114-117, Feb. 2018.
\bibitem{b9} H. Huang et al., “Deep Learning for Physical-Layer 5G Wireless Techniques: Opportunities, Challenges and Solutions,” IEEE Wireless Communications, vol. 27, no. 1, pp. 214-222, Feb. 2020.
\bibitem{b10} Y. Wang et al., “Data-Driven Deep Learning for Automatic Modulation Recognition in Cognitive Radios,” IEEE Transactions on Vehicular Technology, vol. 68, no. 4, pp. 4074-4077, Apr. 2019.
\bibitem{b11} E. Nachmani et al., “RNN decoding of linear block codes,” arXiv preprint arXiv:1702.07560, 2017.
\bibitem{b12} D. E. Rumelhart et al., “Learning internal representations by error propagation,” In Parallel Distributed Processing, vol 1, MIT Press, Cambridge, MA, Sep. 1985.
\bibitem{b13} N. Wu et al., “A CNN-Based End-to-End Learning Framework Toward Intelligent Communication Systems,” IEEE Access, vol. 7, pp. 110197-110204, Jul. 2019.
\bibitem{b14} B. Zhu et al., “Joint Transceiver Optimization for Wireless Communication PHY Using Neural Network,” IEEE Journal on Selected Areas in Communications, vol. 37, no. 6, pp. 1364-1373, Jun. 2019.
\bibitem{b15} Y. LeCun, “Generalization and network design strategies,” Connectionism in perspective, Elsevier, 1989.
\bibitem{b16} T. J. O'Shea, K. Karra, and T. C. Clancy, “Learning to communicate: Channel auto-encoders, domain specific regularizers, and attention,” IEEE International Symposium on Signal Processing and Information Technology (ISSPIT), Limassol, Cyprus, Dec. 2016, pp. 223-228.
\bibitem{b17} V.V. Mannam et.al, “Performance Analysis of Semi-supervised Learning in the Small-data Regime using VAEs”  was published in CSE 40625/60625 ’19: Course Project Workshop, April, 2019, Notre Dame.e, IN .ACM, New York, NY, USA, 4 pages. https://doi.org/10.1145/1122445.1122456
\bibitem{b18} C. Doersch, “Tutorial on Variational Autoencoders,” ArXiv, abs/1606.05908, pp. 1-23, Aug. 2016.
\bibitem{b19} D. P. Kingma and M. Welling, “Auto-Encoding Variational Bayes,” ArXiv, 1312.6114v10, pp. 1-14, May 2014.
\bibitem{b20} T. J. O'Shea et al., “Radio transformer networks: Attention models for learning to synchronize in wireless systems,” 50th Asilomar Conference on Signals, Systems, and Computers, Pacific Grove, CA, Mar. 2016, pp. 662-666.
\bibitem{b21} T. O’Shea and J. Hoydis, “An Introduction to Deep Learning for the Physical Layer,” IEEE Transactions on Cognitive Communications and Networking, vol. 3, no. 4, pp. 563-575, Dec. 2017.
\bibitem{b22} S. Dörner et al., “Deep Learning Based Communication Over the Air,” IEEE Journal of Selected Topics in Signal Processing, vol. 12, no. 1, pp. 132-143, Feb. 2018.
\bibitem{b23} T. Gruber et al., “On deep learning-based channel decoding,” 51st Annual Conference on Information Sciences and Systems (CISS), 2017, pp. 1-6.
\bibitem{b24} H. Kim et al.,“Communication algorithms via deep learning,” arXiv preprint arXiv:1805.09317, 2018.
\bibitem{b25} Y. Jiang et al.,“DEEPTURBO: Deep Turbo Decoder,” IEEE 20th International Workshop on Signal Processing Advances in Wireless Communications (SPAWC), Cannes, France, Jul. 2019, pp. 1-5.


\end{thebibliography}
\end{document}